\documentstyle[preprint,aps]{revtex}
\tightenlines
\begin{document}

\title{Graphical representations and cluster algorithms for
critical points with fields}

\author{O. Redner${}^1$, J. Machta${}^{2}$
\footnote{Permanent address: Department of
Physics and Astronomy, University of Massachusetts, Amherst, MA
01003-3720}, and L. F. Chayes${}^3$}
\address{${}^1$Institut f\"ur Theoretische Physik, Universit\"at
T\"ubingen, Auf der Morgenstelle 14, D-72076 T\"ubingen, Germany\\
${}^2$Department of Physics of Complex Systems,
Weizmann Institute of Science, Rehovot 76100, Israel
\\
${}^3$Department of Mathematics, University of California,
Los Angeles, CA 90095-1555}

\maketitle

\begin{abstract}
A two-replica graphical representation and associated cluster
algorithm is described that is applicable to ferromagnetic Ising
systems with arbitrary fields.  Critical points are associated with
the percolation threshold of the graphical representation.  Results
from numerical simulations of the Ising model in a staggered field
are presented.  For this case, the dynamic exponent for the algorithm
is measured to be less than 0.5.

\end{abstract}
\pacs{05.50.+q,64.60.Fr,75.10.Hk}
\narrowtext

Monte Carlo simulations of equilibrium critical points have been
revolutionized by cluster methods
\cite{WaSw90,SwWa,KaDo,Wolff}.  These methods
effectively reduce
critical slowing for a  broad class of spin models.
However, cluster methods developed
thus far are efficient only for systems
with a high degree of internal or translational symmetry.
In this paper we show how to construct graphical representations
and associated cluster algorithms
appropriate for ferromagnetic Ising systems in the presence of
fields. We demonstrate the method for the Ising system in a
staggered field.

The Swendsen-Wang (SW) cluster algorithm \cite{SwWa} as applied to
the Ising model updates both spin variables and bond variables.
Starting from a spin configuration, satisfied bonds are occupied with
probability $p=1-e^{-2 \beta}$ where $\beta$ is the inverse
temperature.  A bond is {\em satisfied} if the spins at its two ends
agree.  Clusters of spins connected by occupied bonds are
identified.  Each cluster is randomly assigned a new spin value and
each spin in the cluster takes that value.  This constitutes one
Monte Carlo step.  The efficiency of the method is associated with
the fact that the critical point of the Ising model coincides with
the percolation point of the graphical model described by the bond
variables \cite{CoKl,ACCN}. Thus, at criticality, spin-clusters are
coherently updated on all length-scales.

A fundamental problem is encountered in applying the Swendsen-Wang
method when fields are present.  Clusters can be defined in the
usual way but then the fields must accounted for in determining the
probability of flipping  clusters.  Fields may be introduced
directly via a Boltzmann factor \cite{DoSeTa} or via ghost bonds. In
either case, large clusters will tend to be `frozen' in the sense
that they almost always take one spin value.  Because large clusters
are frozen, equilibration occurs slowly on large length scales.  In
addition, the percolation transition of the graphical representation
typically occurs in the disordered phase so that at criticality
there are large clusters that are dynamically frozen.
As a result, the qualitative gains  expected from
cluster methods (characterized by a small value of the dynamic
exponent) are not realized.
Nonetheless, limited quantitative
success has been achieved for the random field Ising model by ad
hoc methods that restrict the cluster size \cite{NeBa}.

In this paper we present a graphical representation and associated
cluster algorithm with the property that the percolation transition
in the graphical representation coincides with the ordering
transition in the spin-system.  Furthermore, at criticality, the
large-scale clusters are free to flip. We achieve this by using a
replica method related to the replica Monte Carlo approaches that
have been applied with some success to spin
glasses~\cite{SwWa86,WaSw88,KaGu}.  We also note a related cluster
method~\cite{DrKr,HeBl96,HeBl98}, in which the system is ``folded''
on itself and pairs of sites are used to make clusters. 

The idea of our
algorithm is as follows.  Two independent replicas of the system in
the same field are simulated simultaneously.  Each site of the
lattice is therefore associated with two spins, one from each
replica, and can be in one of four spin states, $(++)$, $(--)$,
$(+-)$ and $(-+)$.  Sites where the replicas disagree, $(+-)$ and
$(-+)$, are called {\em active} sites. Clusters of active sites are
constructed and flipped. Allowed cluster flips interchange $(+-)$
with $(-+)$. Additional updating is applied independently to each
replica to insure ergodicity. Since there is no net field on active
clusters these flip freely.  It turns out that percolation of active
clusters signals the onset of long range order.

The plan for the remainder of the paper is as follows.  First we
construct a joint distribution of the Edwards-Sokal \cite{EdSo} type
whose spin marginal is two independent Ising models.  Next we
indicate why percolation of the associated graphical representation
coincides with the onset of magnetic ordering in the spin--system. 
We then describe a cluster algorithm which simulates this joint
distribution and present numerical results for the Ising model in a
staggered field.  Finally, we discuss generalizations of the method.

The Hamiltonian for the
Ising model is
\begin{equation}
\label{eqn:ising}
{\cal H}[\sigma]= -\sum_{<i,j>}\sigma_i\sigma_j
 - \sum_i H_i\sigma_i
\end{equation}
where the spin variables,
$\sigma_i$ take the values $\pm 1$.  The first summation is over the
bonds of the
lattice (or more generally, an arbitrary graph).  The
second summation is over the sites of the lattice and the fields
$H_i$ are arbitrary. The Ising model on a square lattice in a
staggered field (equivalently, the Ising antiferromagnet in a uniform
field) is obtained by setting $H_i=+H$ if $i$ is in the even
sublattice and $H_i=-H$ if $i$ is in the odd sublattice.

We now define a joint distribution of two sets of Ising spin
variables, $\{\sigma_i\}$ and $\{\tau_i\}$, and a bond variable
$\{\eta_{ij}\}$. The bond variable is defined for each bond
$<\!i,j\!>$
and takes values $0$ and $1$. We say that $<\!i,j\!>$ is {\em
occupied} if $\eta_{ij}=1$. The statistical weight
$X(\sigma,\tau,\eta)$ for the joint distribution is
\begin{equation} \label{eqn:x} X(\sigma,\tau,\eta)
=e^{-\beta
 \sum_{<i,j>}\sigma_i\tau_i\sigma_j\tau_j +\beta
\sum_i H_i(\sigma_i+\tau_i)}\Delta(\sigma,\tau,\eta) B_{p}(\eta).
\end{equation}
$B$ is the standard Bernoulli factor,
\begin{equation}
B_p(\eta)=p^{|\eta|}(1-p)^{N_b-|\eta|},
\end{equation}
$|\eta|=\# \{<i,j> |\: \eta_{ij}=1\}$  is the number of occupied
bonds and $N_b$ is the total number of bonds of the lattice. The
$\Delta$ factor enforces the rule that only satisfied bonds are
occupied: if for every bond $<\!i,j\!>$ such that $\eta_{ij}=1$ the
spins agree in both replicas ($\sigma_i=\sigma_j$ and
$\tau_i=\tau_j$) then
$\Delta(\sigma,\tau,\eta)=1$; otherwise
$\Delta(\sigma,\tau,\eta)=0$.  Equation (\ref{eqn:x}) is closely
related to the ``red-blue'' graphical representation of the
Ashkin-Teller model given in \cite{ChMa97a}.
It is straightforward to verify that integrating
$X(\sigma,\tau,\eta)$ over $\eta$ yields the statistical
weight for two independent Ising models in the same field,
\begin{equation}
e^{-\beta {\cal H}[\sigma]- \beta {\cal H}[\tau]}=
{\rm const}\sum_{\{\eta\}}
X(\sigma,\tau,\eta)
\end{equation}
if the identification is made that $p=1-e^{-4 \beta }$.

Consider a two-replica spin system in which the $\sigma$-replica
has $(+)$ boundary conditions and the $\tau$-replica has $(-)$
boundary conditions.  The local order parameter is the difference
between the magnetization of the two replicas, $m_i=
(<\!\sigma_i\!>-<\!\tau_i\!>)/2$. Observe that magnetization in a
single Ising model in a field is not generally the correct order
parameter because the field induces local magnetization even in the
disordered phase. By taking the difference between the magnetization
of the two replicas with opposite boundary conditions, this
contribution is canceled leaving only the spontaneous magnetization
induced by the boundary conditions.

Given a bond configuration $\eta$ we can ask for the conditional
probabilities for the spins. Due to the $\Delta$ factor in the
statistical weight, $\sigma_i=+1$ and $\tau_i=-1$ if $i$ is connected
to the boundary by occupied bonds.  On the other hand, due to the
symmetry
of the exponential factor in the statistical weight a
site that is not connected to the boundary is equally likely to
be $(+-)$ or $(-+)$.  Finally, $(++)$ and $(--)$ spin states do not
contribute to $m_i$.  Thus $m_i$ is
{\em exactly}
equal to the probability that $i$ is connected to the boundary and
the onset of long range order coincides with percolation.
For a more detailed argument, see \cite{ChMaRe98b}.

Our replica cluster algorithm simulates two
independent Ising models, $\sigma$ and $\tau$, on the same
lattice and in the same field.
Sites $i$ at which $\sigma_i\neq\tau_i$ are called {\em active}
sites.  Bonds $<\!i,j\!>$ connecting like spins in both
replicas ($\sigma_i=\sigma_j$ and $\tau_i=\tau_j$) are called {\em
satisfied} bonds.

\begin{itemize}
\item
Step 1:  Satisfied
bonds connecting active sites are occupied with probability
$p=1-e^{-4 \beta }$.
\item Step 2:  Clusters of active sites connected by occupied bonds
(including single active sites) are
identified.  The $k^{\rm th}$ cluster is independently assigned a
spin value, $s_k= \pm 1$ with probability $1/2$.  If site $i$ is in
cluster $k$ then the new spin values are
$\sigma_i=s_k$ and $\tau_i=-s_k$.  In this way all active sites are
updated. \item Step 3: Each replica
is independently updated in a way that preserves detailed balance
and insures ergodicity.  This completes one Monte Carlo step.
\end{itemize}

Without Step 3, the algorithm is not ergodic since the product
$\sigma_i\tau_i$
is locally conserved.  Step 3 can be
implemented in many ways.  For example, each replica can be
separately updated using the Metropolis algorithm.  For the
staggered field model in periodic boundary conditions we can effect
further mixing by translating the $\tau$ replica by a random amount
relative to the $\sigma$ replica. If the translation is an odd
vector, all $\tau$ spins are flipped. Since the Hamiltonian is
invariant with respect to even translations and odd translations
plus spin flips it is clear that the translation part of the
algorithm satisfies detailed balance.
(Note that Metropolis sweeps
are required even with random translations because the net
staggered
magnetization,  $s=[\sum_{i \: \varepsilon \: {\rm odd}}-\sum_{i \:
\varepsilon \: {\rm even}}](\sigma_i + \tau_i)$ is otherwise a
conserved quantity.)
The simulations reported  here implement Step 3 with both a 
Metropolis sweep and a random translation.

The validity of the algorithm is proved by showing that it is
ergodic and that the joint distribution, $X(\sigma,\tau,\eta)$,
 defined in Eq.\
(\ref{eqn:x}) is the stationary distribution of the algorithm.
Ergodicity follows immediately from Step 3.  To prove stationarity
we observe, following \cite{EdSo}, that the
Steps 1 and 2 of the algorithm correspond to conditional
probabilities associated with $X(\sigma,\tau,\eta)$. Step
1 is the conditional probability of a bond configuration given a spin
configuration.  Note that the bonds connecting inactive sites are
not actually specified in the algorithm but since bonds are
independently occupied this is of no consequence. Step 2 is the
conditional probability of the spin configuration on the active
sites given a bond configuration, a set of active sites and the spin
configuration on the inactive sites.

We simulated the square lattice, staggered field Ising model in
periodic boundary conditions using the two-replica cluster algorithm
described above.  Data was collected for $\beta H=0$, $2$ and $4$
and for size $L$ in the range 16 to 256.  Each $L$ and $\beta
H$ was simulated for 50,000 Monte Carlo steps, dropping the first
5,000.  Figure \ref{fig:1} shows the probability that there is a
spanning cluster ${\cal S}$ as a function of temperature $T$ for
several system sizes and $\beta H=2$. Spanning is defined as
wrapping around the torus in either direction. The vertical line is
the high precision value of $T_c$ given in \cite{BlWu}.  Figure
\ref{fig:2} is a plot of ${\cal S}$ versus $c(\beta H) [T-T_c(\beta
H)] L$ of the data for all values of $\beta H$ and $L$. Data collapse
is achieved using $T_c(\beta H)$ from \cite{BlWu}, $c(0)=1$,
$c(2)=2.64$ and $c(4)=7.30$.  This figure illustrates that the model
is in the Ising universality class independent of $\beta H$. The fact
that ${\cal S} \rightarrow 1/2$ for large systems and $T<T_c$ is due
to periodic boundary conditions. For half of the Monte Carlo steps,
replicas are magnetized in the same direction preventing active
clusters from spanning.  These results provide a clear numerical
verification that percolation in the two-replica
representation is coincident with the critical point.

The Table shows the integrated autocorrelation time for
the absolute value of the magnetization of one replica, $\tau_m$ and
the net staggered magnetization $\tau_s$ versus system size.  The
integrated autocorrelation time is
$1/2$ plus the sum of the
normalized autocorrelation function from time 1 through 200. The
system size dependence of $\tau_s$ and $\tau_m$ can be reasonably fit
either as $A L^z$ or as $A + B\log(L)$.  For the whole range of $L$,
logarithmic growth gives a better fit visually.  Fitting a power law
for system sizes greater than $40$, we find dynamic exponents,
$z_m=0.19 \pm .09$ and $z_s=0.33 \pm .09$ for $\beta H=4$ with
nearly identical results for the other two field values. The quoted
error is the statistical part and does not include systematic errors
due to finite system size and finite cut-off in summing the
autocorrelation function.  A conservative conclusion is that
$0(\log) \leq z < 0.5$.  It is clear that the algorithm achieves
considerable acceleration over local dynamics, where $z \geq 2$.
This is perhaps surprising in view of the fact that the algorithm
uses local dynamic to break conservation of staggered
magnetization.  The autocorrelation times for the present algorithm
and the ordinary Ising model ($\beta H=0$) are roughly a factor of
five larger than for the Swendsen-Wang algorithm however, further
study is needed to determine whether the two algorithms share the
same $z$.  On the other hand, in the presence of a staggered field,
autocorrelation times for both the Swendsen-Wang algorithm and the
two-replica algorithm without translations are much larger than the
values obtained here.  Rough estimates of exponential
autocorrelation times for the present algorithm show that they are
about twice the corresponding integrated autocorrelation time.

In the case of a staggered field, we
have made use of the symmetries of the problem to
incorporate as much mixing as possible into Step 3 of the
algorithm.  For systems such as the random field Ising model which
do not enjoy translational symmetries, local dynamics
are all that is available to equilibrate the average magnetization
at each site.  Thus, the two replica algorithm may not be a
qualitative improvement over local dynamics alone.  However,
significant acceleration may be achieved by using many replicas.
Suppose we have $2K$ replicas, $\{\sigma^{(l)}
| l=1,\ldots, K\}$ all in the same field $\{H_i\}$. In each
Monte Carlo step,  the replicas are randomly paired and the two
replica cluster procedure is applied to each pair.  Each replica is
also updated independently by some local ergodic algorithm. The
replica summed magnetization at each site, $\sum_{l=1}^{2K}
\sigma^{(l)}_i$, is conserved except by the local dynamics and so may
relax slowly to equilibrium.  This implies that the average
magnetization at any site may reach equilibrium slowly resulting in
a large exponential autocorrelation time.  However, once the
equilibrium values of the magnetization are reached,  the
fluctuations of the replica summed magnetization are small for large
$K$ and thus couple weakly to the observables of individual
replicas. This may yield rapid decorrelation and small values of
integrated autocorrelation times.  Note that in the two replica
simulations, we observed $\tau_{\rm exp}$ to be about twice
$\tau_{\rm int}$.

We thank E. Domany for important contributions in the early stages
of this work and D. Kandel for useful discussions. J. M. gratefully
acknowledges receipt of the Michael Visiting Professorship of the
Weizmann Institute of Science during the period that this work was
carried out.  We thank the Institut f\"ur Theoretische Physik of the
Universit\"at T\"ubingen for use of their computational facilities.
This work was also supported by National Science Foundation grant
DMR-9632898.

\newpage
\begin{table}
\caption{Integrated autocorrelation times for the absolute value of
the magnetization of a single replica $\tau_m$ and the net staggered
magnetization of both replicas $\tau_s$.  For each entry,
the one standard deviation error is 13\%.}
\begin{tabular}{l||l|ll|ll}
 & \multicolumn{1}{c|}{$\beta H=0$} & \multicolumn{2}{c|}{$\beta
H=2$} &
   \multicolumn{2}{c}{$\beta H=4$} \\
$L$ & $\tau_m$ & $\tau_m$ & $\tau_{s}$ & $\tau_m$ & $\tau_{s}$
\\
\hline
16 &  9.3 & 10.7 &  8.1 & 17.1 & 15.0 \\
24 & 14.5 & 14.8 & 12.6 & 22.2 & 19.9 \\
32 & 17.5 & 20.4 & 15.6 & 28.2 & 23.6 \\
40 & 22.1 & 27.2 & 23.9 & 22.4 & 21.0 \\
48 & 28.2 & 31.3 & 26.6 & 31.5 & 29.4 \\
56 & 29.6 & 27.8 & 22.6 & 36.1 & 34.3 \\
64 & 28.6 & 34.0 & 30.6 & 30.4 & 27.3 \\
80 & 32.5 & 34.9 & 35.6 & 30.2 & 29.8 \\
96 & 35.5 & 36.2 & 31.0 & 36.2 & 37.2 \\
112 & 29.6 & 38.6 & 36.4 & 38.8 & 43.0 \\
128 & 30.0 & 35.9 & 35.9 & 36.7 & 40.1 \\
144 & 35.9 & 37.4 & 37.5 & 37.7 & 43.0 \\
160 & 35.3 & 37.5 & 35.0 & 40.9 & 47.6 \\
192 & 42.6 & 37.0 & 36.7 & 42.9 & 46.3 \\
256 & 37.8 & 44.4 & 48.4 & 40.1 & 46.9
\end{tabular}
\end{table}

\begin{figure}
\caption{The spanning probability vs.\
$T$ for various system sizes and $\beta H=2$.}
\label{fig:1} \end{figure}

\begin{figure}
\caption{The spanning probability ${\cal S}$ vs.\
$c(\beta H) [T-T_c(\beta H)] L$ for all system sizes and the three
values of $\beta H$.}  \label{fig:2}
\end{figure}

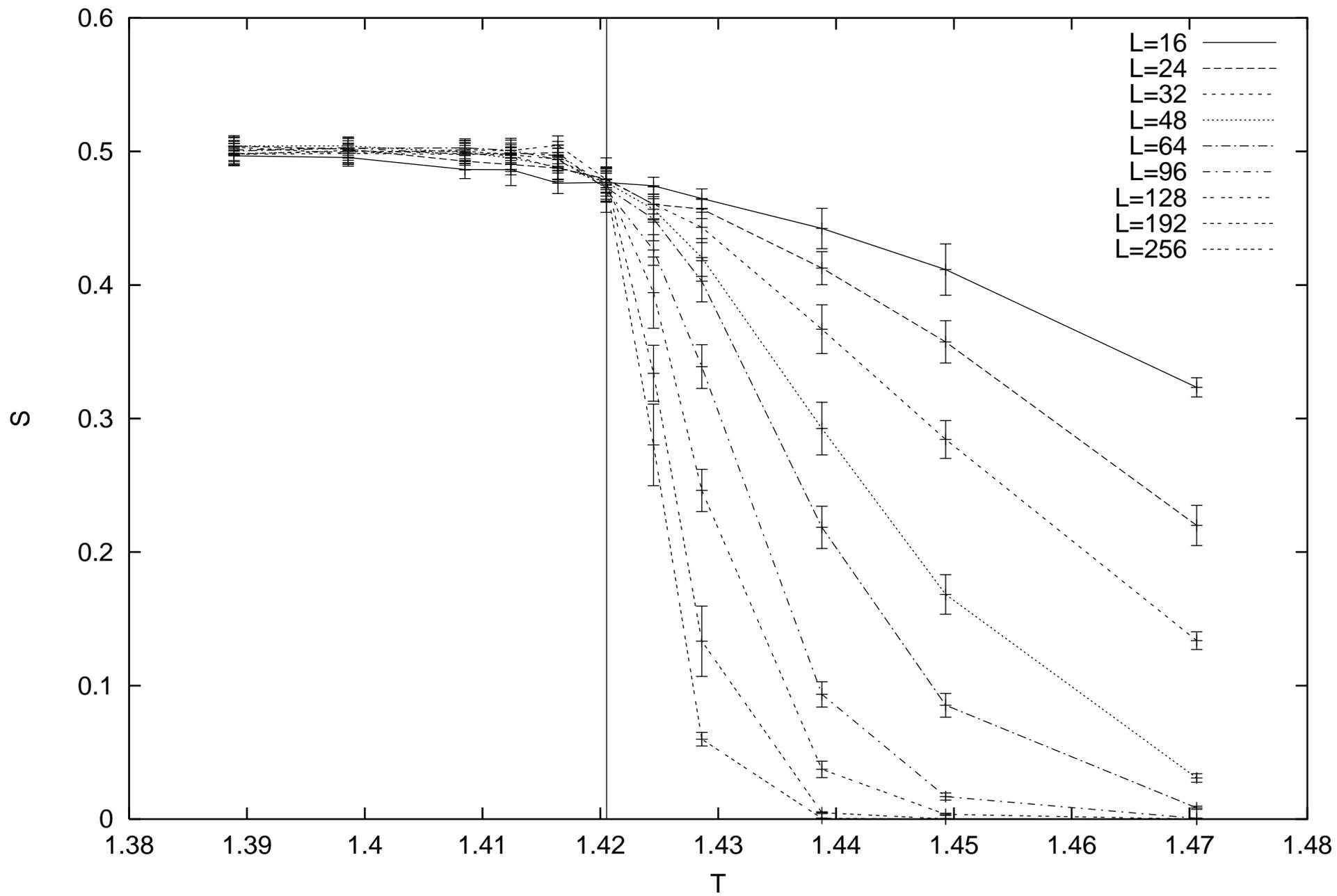
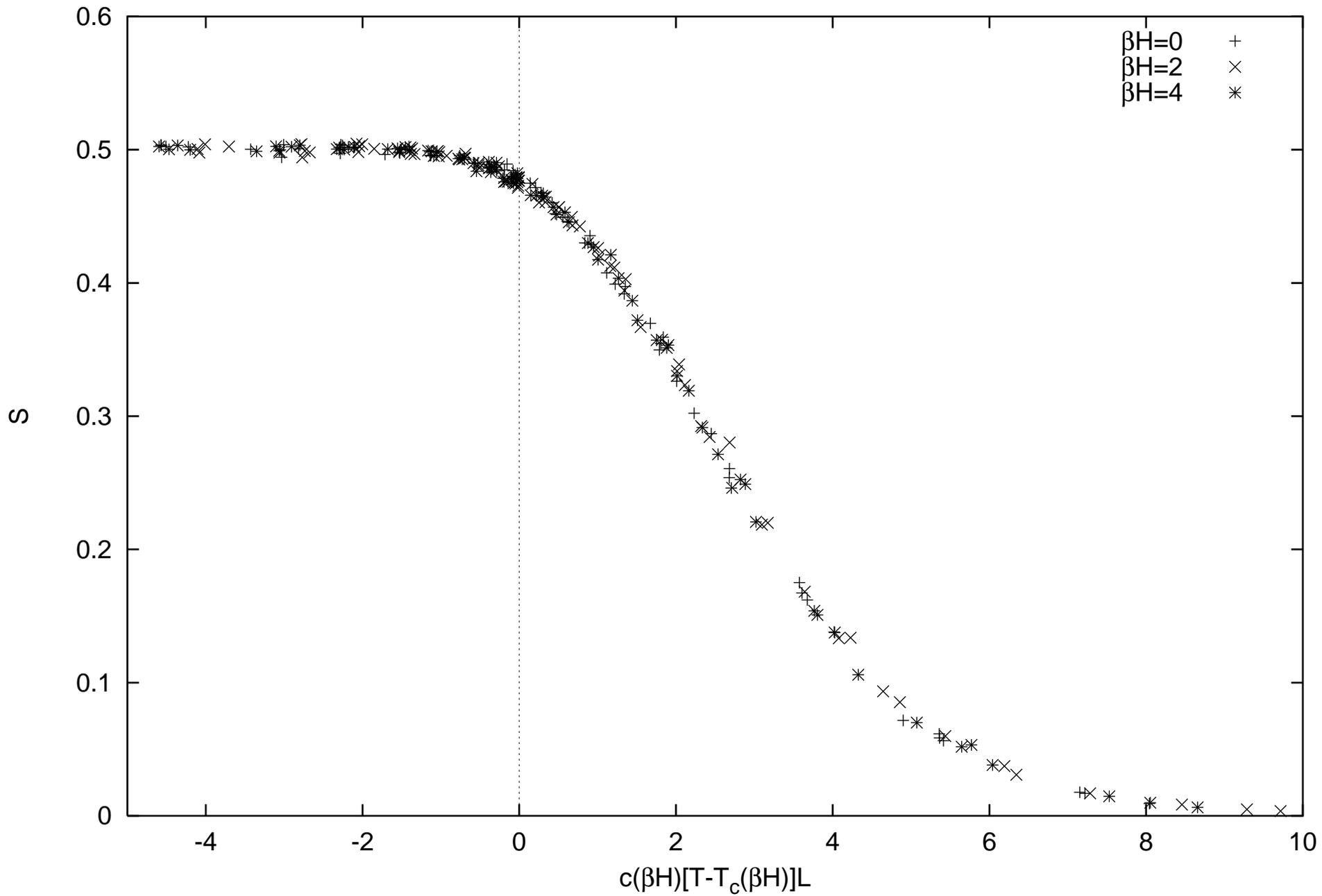
\end{document}